# Invisible Labor, Visible Barriers: The Socioeconomic Realities of Women's Work in Pakistan

Sana Khalil[1] and Angela Warner[2]

[1] University of Washington (Tacoma), 1900 Commerce St, Tacoma, WA 98402, USA
[2] University of Washington (Tacoma), 1900 Commerce St, Tacoma, WA 98402, USA

**Abstract.** We provide a brief note highlighting various barriers that shape women's economic opportunities in Pakistan, where female labor force participation remains among the lowest globally. Labor force surveys (2020–21) show a stark rural–urban divide with rural women's participation at 28% compared to 69% for rural men, while urban women's participation is 10% compared to 66% for urban men. Unemployment is higher for women in both rural (7%) and urban (16%) areas than for men (5% and 6%, respectively). Women are concentrated in agriculture (68%), with limited presence in services (17%) and industry (15%), and largely in rural (51%) or home-based (30%) work. Only 14% work in formal business settings. Employment status reflects high vulnerability, with 63% of rural women as unpaid contributing family workers versus 17% of urban women. Interviews with women in Karachi highlight childcare constraints, harassment and safety concerns, transport barriers, and family opposition as key impediments.

**Keywords:** Female labor force participation, Gender norms, Intrahousehold dynamics, Discrimination.

## 1 Introduction

Pakistan's labor market exhibits some of the starkest gender disparities in the world. The Global Gender Gap Report (2024) ranks Pakistan 145th of 146 with an overall parity score of 0.570, reflecting especially large deficits in economic participation and political empowerment. Although women's literacy rates and enrollment in higher education have improved over the past two decades, their access to jobs in both the public and private sectors remains limited. As of 2023, only 24% of women aged 15+ participate in the labor force (modeled ILO estimate; International Labour Organization, 2025), far below the 32% average for working-age women in South Asia (World Bank, 2024). According to the Pakistan Labor Force Survey 2020-21, women hold only 5.7% of managerial roles, while among college-educated workers, unemployment is 43% for women versus 7% for men (Pakistan Bureau of Statistics, 2025).

Gallup Pakistan's analysis of the 2017–18 Pakistan Labor Force Survey also highlights wide gender gaps. Among individuals aged 18+ with at least an undergraduate degree, 41% of women were unemployed versus 6% of men. In the same group, the labor force participation rate was 48% for women and 94% for men. Nearly 83% of unemployed women in this cohort were willing to accept jobs with lower terms and conditions, compared to 59% of unemployed men.

Existing research on female labor force participation (FLFP) in Pakistan focuses on individual and household characteristics such as education, marital status, and fertility (Shah, 2004; Faridi and Rashid, 2014), as well as intra-household power structures and marriage institutions (Khalil, 2021; 2024). Women in close-kin marriages and extended family systems often face increased obligations in hosting, caregiving, and household management—activities that serve as markers of social prestige but divert time from paid work, frequently confining them to home-based informal employment (Akram-Lodhi, 1996; Kaplan et al., 2016; Islam et al., 2018; Turaeva and Becker, 2022).

Survey evidence from Pakistan reveals deep-seated gender biases. For example, a nationally representative poll found that 85% of Pakistanis agree that "when jobs are scarce, men should have more right to a job than women," compared to 33% globally and 50% who disagreed (World Values Survey; Gallup and Gilani Pakistan Poll, 2017–2019).

On the demand side, explicit discriminatory hiring remains widespread. Employers often state gender preferences in job ads, with women less likely to be preferred in most occupational categories (Khalil, 2024). Matsuda et al. (2019), analyzing postings from Pakistan's largest employment portal, found that listings favor men for higher-level positions requiring professional experience, while women are more often sought for roles requiring higher education but remain underrepresented in managerial jobs. Paradoxically, 67% of respondents in a Gallup poll said gender equality had been achieved at their workplace, suggesting normalization of constraints rather than their absence (Gallup and Gilani Pakistan and WIN World Survey, 2020–21).

This brief note summarizes various factors that limit Pakistani women's participation in paid work. Using national statistics (Pakistan Labor Force Surveys, ILO, World Bank) and selected literature, it documents the rural–urban divide, pronounced occupational segregation (with 68% of employed women in agriculture), the prevalence of unpaid or home-based work, and women's marginal presence in formal workplaces. It also draws on qualitative surveys from married women in Karachi, Pakistan, to highlight barriers such as mobility and safety concerns, transportation and childcare constraints, and family opposition to participation in paid work outside the home.

## 2 Gendered Labor Market Outcomes

### 2.1 Rural-Urban Divide

Data from the Pakistan Labor Force Survey (PLFS) 2020–21 reveals persistent gender gaps and a marked rural–urban divide (Table 1). Across both urban and rural areas, women's participation rates are consistently lower than men's. The gap is particularly stark in urban areas, where in 2020–21, men participated in the labor force at a rate approximately 6.6 times higher than women (65.9% vs. 10%). While rural women perform slightly better with a participation rate of 28%, they still trail far behind rural men, who participate at a rate of 69.1%.

A decade-long comparison from 2010–11 to 2020–21 suggests stagnation rather than structural progress. During this period, rural women's labor force participation increased by just 0.7 percentage points, while urban women's participation declined by 0.8 percentage points. Men's participation rates also declined slightly in both areas, resulting in a marginal narrowing of the gender gap—not due to improved female participation, but because of falling male rates.

**Table 1.** Men and women's work trends in Pakistan, 2010-2021.

| **Indicator** | **2010-11** | | **2020-21** | |
|---|---|---|---|---|
| | **Men** | **Women** | **Men** | **Women** |
| Rural labor participation rate | 64.5 | 27.3 | 69.1 | 28 |
| Urban labor participation rate | 80.5 | 10.8 | 65.9 | 10 |
| **Refined [1]Activity Rate** | **68.7** | **21.7** | **67.9** | **21.4** |
| Rural unemployment rate | 4 | 6.4 | 5.1 | 7.4 |
| Urban unemployment rate | 7.1 | 20.7 | 6 | 16.4 |
| **Total unemployment rate** | **5.1** | **8.9** | **5.5** | **9** |

Pakistan Labor Force Survey, 2020-2021

Unemployment data from PLFS 2020-21 also indicate a double disadvantage for women: not only are they less likely to participate in the labor force, but those who do are more likely to be unemployed. This issue is especially pronounced in urban areas, where women face an unemployment rate nearly three times that of men (16.4% vs. 6%). Even in rural areas, women's unemployment remains higher (7.4% vs. 5.1%). Over the decade, unemployment among urban women decreased by 4.3 percentage points and among urban men by 1.1 points; however, rural unemployment worsened for both genders.

The survey also reveals large differences in the types of employment across gender and place of work. Rural men are more likely to be own-account workers (44.7%) compared to urban men (33%) and are also more likely to work as contributing family workers[2] (12.9% vs. 7.2%). For women, the contrast is more pronounced: rural women are overwhelmingly engaged in contributing family work (63.2%) compared to only 17.2% of urban women. Own-account work is similar for both groups (18.9% for rural and 19.5% for urban women). Urban women, however, are significantly more likely to be employees (63%) than rural women (17.8%).

[1] Refined activity rate refers to the proportion of the currently active population in the total population aged 10 years and above.

[2] According to the International Labour Organization (ILO), contributing family work refers to market-oriented tasks performed for family members, which may be compensated through cash, in-kind payments, or remain unpaid. Such compensation is typically irregular and may include non-monetary benefits. Among Asian countries with comparable cultural and social norms, Pakistan reports the highest proportion of female contributing family workers—significantly higher than India (29%) and Bangladesh (27%).

On an international scale, Pakistan stands out for having the highest share of female contributing family workers in South Asia. According to World Bank/ILOSTAT modeled estimates for 2023, nearly 53% of employed women in Pakistan are contributing family workers, compared to 35% in India and 27.9% in Bangladesh (World Bank/ILO, 2025). Contributing family workers are engaged in a relative's market-oriented enterprise—such as a family business or farm—typically with little or no direct wages. As a result, women in these roles remain in particularly vulnerable economic positions. Although they are counted as part of the labor force, this work often goes unrecognized due to the absence of formal records, and many women may not even report such activities as employment because of prevailing social stigma around women's participation in paid work.

Earlier Pakistan Labor Force Surveys also indicate a high incidence of women's unpaid contributing family work. More than half of employed women—56% in 2014–15 and 57% in 2018–19—were engaged as unpaid or contributing family workers (Andlib and Zafar, 2023). Andlib and Zafar's analysis shows that young girls aged 10–14 are disproportionately represented in this category, and that factors such as large household size, rural residence, and a male-headed household working in agriculture significantly increase the likelihood of women becoming contributing family workers. By contrast, women with higher levels of education and those with migration experience are less likely to fall into this form of unpaid labor. The normalization of unpaid work within patriarchal structures not only reinforces women's economic dependence but also limits their access to formal employment, decision-making power, and broader empowerment opportunities. Women's work in Pakistan, therefore, must be understood within the broader social structures and constraints that shape it. In this regard, Stier and Mandel (2009) emphasize that gender inequality in labor markets cannot be separated from family structures and institutional contexts. In countries with more egalitarian family arrangements and labor market institutions, women's earnings play a stronger role in reducing household inequality. By contrast, in more traditional or weakly regulated labor markets, women's economic contributions often remain marginal or secondary, limiting their potential equalizing effect.

More recent data confirm that these inequalities extend across employment sectors. The 2020–21 Labor Force Survey shows that agriculture overwhelmingly dominates women's employment, absorbing 68% of female workers compared to only 28% of men. Men, by contrast, are more diversified, with 43% employed in services and 29% in industry. Women's participation in these potentially higher-paying sectors is much lower (17% in services and 15% in industry). Comparing South Asian economies, World Bank (2022) data show female employment in services is highest in Sri Lanka (47%), followed by Bangladesh (27%), India (25%), and lowest in Pakistan (17%). These patterns reflect occupational segregation, which likely contributes to Pakistan's substantial gender wage gap.

There are stark gender differences in work locations as well. Most women are employed in rural settings (51%) or in home-based work (30%), accounting for over 80% of all female employment. In contrast, men's employment is more varied: nearly half (48%) work in formal business settings such as shops, offices, or industrial facilities, while just 1.5% work from home.

While rural employment is significant for both genders, it is substantially higher for women (51%) than for men (30%). Informal workspaces, such as employers' homes, streets, or undefined locations, show a clear gender divide, with more men (20.5%) working in these environments than women (5%).

A number of factors explain these inequalities, including societal norms, kinship systems, and gender roles. For example, the ideals of purdah (female seclusion) and women as keepers of family "honor" can reinforce preferences for domestic roles and limit women's mobility and willingness to engage in paid work outside the home (Khalil, 2022; 2024).

Findings from qualitative surveys (discussed in the next section) suggest that many women working in the informal sector perceive paid work outside the home as a last resort, driven by poverty, lack of resources, or the absence of male earners. Most respondents expressed a preference for staying within the "respectable" domestic sphere (*chadar and chardewari*) and favored home-based or female-dominated occupations (Heintz et al., 2018). This also aligns with Dildar's (2015) findings from the 2008 Turkish Demographic Health Survey, which show that religious practices significantly lower women's labor force participation, especially in urban areas. Dildar found that having an additional child under the age of 5 reduced labor force participation by 7.4% for urban women and 3.2% for rural women. She attributes this gap to stronger patriarchal constraints in urban areas, while rural women often work on family farms, paid or unpaid.

Several factors may explain the urban–rural gap in women's labor force participation. In urban settings, higher male earnings may reduce the perceived necessity of a second income, especially where the male breadwinner model prevails and women are viewed as secondary earners. Conversely, rural women may find it easier to combine agricultural work with caregiving responsibilities, compared to the rigid time structures of urban jobs. Additionally, gender bias in hiring may be more prevalent in urban labor markets, where employers have access to a broader pool of candidates and may engage in more discriminatory hiring practices (Khalil, 2021).

## 3 Selected Literature: Barriers to Women's Paid Work in Pakistan

### 3.1 Intrahousehold Dynamics

The extended family system plays a key role in women's autonomy and employment. Ahmad et al. (2016) find a strong negative relationship between joint family structure and women's empowerment, with larger households linked to reduced decision-making power. Within households, women's position depends on age, relation to core members, and having sons. Extended families often oppose women's higher education, outside employment, or academic careers (Fazal et al., 2019).

In Pakistan, marriage, women's work, and domestic care are closely connected. Some studies suggest that married women have more mobility (Shah et al., 1976; Faridi et al., 2009) and higher paid labor participation than unmarried women (Ejaz, 2007; Faridi and Rashid, 2014), yet joint families increase unpaid domestic burdens (Khalil, 2021, 2024). Family size, assets, presence of children, and joint living arrangements correlate negatively with women's workforce participation (Ejaz, 2007). Rural women are more often employed, partly due to home-based production opportunities (Faridi and Rashid, 2014). Women, even when employed outside the home, retain primary domestic roles (Ahmed, 2020) and bear a disproportionate work–family burden that hinders career advancement (Sarwar and Imran, 2019). Barriers to employment also include stigma around women's mobility and interactions with non-kin men (Ahmed, 2020).

Kin marriages can also influence women's bargaining power within the household through the consolidation of assets within the patriline. Where a woman's inheritance is appropriated informally by her husband or in-laws, her fallback position weakens, reducing her leverage in decision-making; conversely, retaining control over her inheritance can enhance bargaining power (Bahrami-Rad, 2021). The physical proximity between natal and affinal kin in endogamous unions can intensify surveillance aimed at enforcing ideals of "purity," thereby restricting women's interactions with non-kin men and steering their occupational opportunities toward home-based work (Jayachandran, 2021).

### 3.2 Gendered Barriers in the Workplace

Demand-side discrimination plays a crucial role in limiting women's opportunities in the labor market, contributing to persistent gender inequality in labor market outcomes. In Pakistan, it is common for employers to explicitly state gender preferences in job advertisements. These ads often reinforce gendered expectations, with women being preferred for a narrow range of roles, such as teaching and healthcare, while being excluded or less favored for most other occupations (Khalil, 2021). This practice not only discourages women from pursuing certain career paths but also perpetuates job segregation. Nasir (2005) argues that, in Pakistan, employers often categorize jobs as "men's" or "women's" based on traditional gender norms, rather than the specific qualifications required for the role.

The literature documents significant gender segregation and bias in employment opportunities. Sarwar and Imran (2019) note that certain jobs are perceived as suitable only for men, with women typically confined to sectors like healthcare and teaching. Gender bias is also reflected in hiring and promotion practices that benefit men. Zulfiqar et al. (2024) highlight that hiring staff may refuse to consider women for promotions if they have children, due to the assumption that they will be too busy to adequately fulfill job requirements. Sarwar and Imran (2019) further observe that human resource managers typically prefer promoting men, and that women face greater difficulty securing interviews, receiving job offers, being considered for promotions, and being invited to training. Similarly, Waqar et al. (2021) find that the relationship between work experience and number of promotions differs significantly by gender: men tend to receive more promotions as their experience increases, while women may go their entire careers without being promoted.

Gender-based wage disparities in Pakistan are frequently examined using non-experimental methods such as wage regression analysis. These studies consistently reveal a significant unexplained wage gap, often attributed to discrimination. For instance, Siddiqui et al. (1998) and Sabir and Aftab (2007) find that even after controlling for education, experience, and job type, a substantial portion of the wage gap remains unexplained. Sabir and Aftab, in particular, note an increase in gender discrimination after 2000, especially at lower wage quantiles, suggesting that employer bias has played a significant role in widening wage disparities.

Workplace harassment constitutes another major barrier to women's employment. The literature consistently finds sexual harassment to be widespread, with women sometimes deterred from working, fired for making allegations, pressured to ignore harassment, or even coerced into tolerating it to advance their careers (Zulfiqar et al., 2024; Sarwar and Imran, 2019).

Structural and resource limitations further complicate women's employment situations. A lack of suitable transportation poses a significant challenge. Fazal et al. (2019) note that distant workplaces without residential accommodation necessitate long commutes, which are often unfeasible for women balancing domestic responsibilities, thereby affecting work performance. Mobility restrictions—where women may require permission from husbands, male family members, or mothers-in-law to travel, compound these challenges. Insufficient workplace support for working mothers adds another layer, with Sarwar and Imran (2019) identifying the lack of childcare facilities, inadequate maternity leave, inflexible working hours, and limited remote work opportunities as key barriers to women's employment and career advancement after motherhood.

## 4 Qualitative Research

To better understand the social, cultural, and economic factors influencing married women's paid work in Pakistan, a qualitative survey of married women was carried out in the city of Karachi, the capital of Sindh province, from August 2019 to September 2022. [3] In total, 50 structured interviews were conducted. Participants were recruited using snowball sampling[4], an approach suited to contexts where social norms discourage discussion of marital life. Respondents varied in ethnicity, education, household income, and employment: some were paid domestic workers with little formal schooling, others were professionals (e.g., teachers, software engineers, customer-service staff). Several were seasonal migrants from southern Punjab working in Karachi as domestic help and returning home for sowing/harvest seasons; most preferred domestic care work in Karachi over farm work due to better access to health care, education, and distance from village-level stigma around women's paid work.

[3] Due to the COVID-19 lockdowns, interviews were conducted in two phases. The first phase of surveys was conducted from August 2019 to December 2019, followed by a second phase from January 2022 to September 2022.

[4] The sample is non-random and therefore is not representative of the entire population of married women in Pakistan or the city of Karachi. However, the findings are intended to offer qualitative insights to understand various intrahousehold dynamics concerning women's paid work in Pakistan.

To ensure trust and protect participants' privacy, respected community figures, such as local elders, were engaged as liaisons. The questionnaire was provided in Urdu, with peer translators assisting respondents who spoke local dialects. Observing that women were more open with interviewers from their communities, interviews were arranged accordingly. As the sample was selected through non-random snowball sampling, it is not representative of all women in Pakistan. To accommodate nuanced responses, the survey included multiple-response questions, so some categories may exceed 100% in reported percentages.

The qualitative survey results reveal multiple, interconnected barriers influencing married women's labor market participation in Karachi. Women's household decision-making and autonomy remain heavily shaped by patriarchal norms. While 45% of women reported sharing financial decisions with their husbands, 30% said their husbands made all such decisions, and only 15% had full control themselves. Mobility is similarly restricted, where over 40% reported facing severe limitations on their movement, and only 20% enjoy minimal restrictions. Harassment fears prevent 68% from traveling alone after dark, and 52% require permission to visit friends or leave the house.[5] These constraints severely limit access to employment and social engagement. Marriage patterns further influence autonomy, with 90% in arranged marriages—60% within kinship networks—where family often controls both mobility and work choices. In terms of employment, 55% of women prefer home-based work due to flexibility and alignment with family expectations. However, significant barriers persist, including childcare responsibilities (45%), safety concerns (40%), family opposition (38%), transportation difficulties (32%), and social stigma (35%). When asked about preferred sectors for paid work, many participants gravitate toward roles deemed culturally acceptable and safe, such as teaching in girls' schools (40%) or healthcare roles in women's clinics (25%). Home-based businesses appeal to 20%, while fewer prefer corporate (10%) or factory work (5%), citing mixed-gender environments and safety risks. Broader, overlapping barriers include harassment (75%), transportation (65%), family opposition (55%), social stigma (60%), and work–family balance (70%), suggesting the complex web of factors that shape women's economic participation.

Overall, the findings of the qualitative survey pinpoint the various barriers that married women in Karachi face in entering and advancing within the labor market. These constraints are embedded in intersecting cultural norms, patriarchal household structures, and gendered expectations regarding work and family roles, patterns consistent with broader evidence from Pakistan and comparable contexts (Sarwar and Imran, 2019; Zulfiqar et al., 2024; Khalil, 2021). Mobility restrictions, often enforced through practices such as requiring permission to travel or prohibiting movement after dark, mirror national trends in which safety concerns, harassment fears, and purdah norms limit women's engagement in public spaces (Jayachandran, 2021; Ahmed, 2020). Limited decision-making authority within households, particularly in financial matters, reflects entrenched patriarchal arrangements documented in both quantitative and qualitative research (Ahmad et al., 2016; Habiba et al., 2016; Khalil, 2024).

[5] The percentages don't add to 100% because the responses are based on multiple-response questions.

## 5 Conclusion

Women's work in Pakistan is shaped by a web of structural, cultural, and institutional constraints that systematically limit their access to decent employment. Pakistan Labor Force Surveys indicate persistent gender gaps in labor force participation, unemployment, and occupational distribution, with women disproportionately concentrated in agriculture and unpaid family labor, while remaining underrepresented in services, industry, and formal workplaces. Comparative evidence across South Asia highlights Pakistan's particularly high incidence of contributing family work among women, reflecting the country's entrenched reliance on vulnerable and informal labor.

Micro-level insights from Karachi further reveal how household decision-making, mobility restrictions, childcare responsibilities, harassment fears, and social stigma intersect to confine women to home-based or culturally "acceptable" occupations. These patterns mirror findings from earlier studies showing how patriarchal kinship systems, joint family structures, and discriminatory hiring practices reinforce women's economic dependence and marginalization. Taken together, the evidence shows that barriers to women's labor force participation in Pakistan are not isolated but embedded in overlapping economic, social, and cultural structures.

Addressing these challenges requires not only economic reforms but also broader shifts in gender norms, workplace practices, and family structures to enable more inclusive and equitable labor market outcomes. Legal protections against workplace harassment and discrimination must be accompanied by investments in childcare, transport, and safe public spaces to enable women to enter and remain in the workforce.

### References

1. Ahmad, N., et al.: Gender equality and women's empowerment in rural Pakistan. In: Agriculture and the Rural Economy in Pakistan: Issues, Outlooks, and Policy Priorities, pp. 391–432. University of Pennsylvania Press on behalf of the International Food Policy Research Institute (IFPRI), Philadelphia (2016).
2. Ahmed, S.: Women left behind: Migration, agency, and the Pakistani woman. Gender and Society 34(4), 597–619 (2020).
3. Akram-Lodhi, A.H.: "You are not excused from cooking": Peasants and the gender division of labor in Pakistan. Feminist Economics 2(2), 87–105 (1996).
4. Andlib, Z., Zafar, S.: Women and vulnerable employment in the developing world: Evidence from Pakistan. Journal of International Women's Studies 25(7), 17 (2023).
5. Bahrami-Rad, D.: Keeping it in the family: Female inheritance, inmarriage, and the status of women. Journal of Development Economics 153, 102714 (2021).
6. Dildar, Y.: Patriarchal norms, religion, and female labor supply: Evidence from Turkey. World Development 76, 40–61 (2015).

7. Economic and Social Commission for Asia and the Pacific: Demographic changes. Available at: https://www.population-trends-asiapacific.org/data/PAK (2023).
8. Ejaz, M.: Determinants of female labor force participation in Pakistan: An empirical analysis of PSLM (2004-05) micro data. Lahore Journal of Economics 12(Special Edition), 204–233 (2007).
9. Faridi, M.Z., Chaudry, I.S., Anwar, M.: The socio-economic and demographic determinants of women's work participation in Pakistan: Evidence from Bahawalpur District. South Asian Studies, 353–369 (2009).
10. Faridi, M.Z., Rashid, A.: The correlates of educated women's labor force participation in Pakistan: A micro-study. Lahore Journal of Economics 19(2) (2014).
11. Fazal, S., et al.: Barriers and enablers of women's academic careers in Pakistan. Asian Journal of Women's Studies 25(2), 217–238 (2019).
12. Gallup and Gilani Pakistan: 40 years of polling on women and gender related issues in Pakistan (1979–2019). Available at: https://gallup.com.pk/wp/wp-content/uploads/2019/07/Women-Polling-report-3.pdf (2019).
13. Gallup Pakistan: Survey findings. Available at: https://gallup.com.pk/post/30985 (2021).
14. Habiba, U., Ali, R., Ashfaq, A.: From patriarchy to neopatriarchy: Experiences of women from Pakistan. International Journal of Humanities and Social Science 6(3), 212–221 (2016).
15. Heintz, J., Kabeer, N. and Mahmud, S.: Cultural norms, economic incentives and women's labour market behaviour: empirical insights from Bangladesh. Oxford Development Studies 46(2), 266–289 (2018).
16. Islam, M.M., Ababneh, F.M., Khan, M.H.R.: Consanguineous marriage in Jordan: An update. Journal of Biosocial Science 50(4), 573–578 (2018).
17. International Labour Organization: Global Wage Report 2018/19: What lies behind gender pay gaps. International Labour Office – Geneva: ILO (2018).
18. International Labour Organization: ILO Modelled Estimates and Projections database (ILOEST). Available at: https://ilostat.ilo.org/data (accessed 28 March 2025).
19. Jayachandran, S.: Social norms as a barrier to women's employment in developing countries. IMF Economic Review 69(3), 576–595 (2021).
20. Kaplan, S., et al.: The prevalence of consanguineous marriages and affecting factors in Turkey: A national survey. Journal of Biosocial Science 48(5), 616–630 (2016).
21. Khalil, S.: Structures of constraint and women's paid work in Pakistan. Review of Socio-economic Perspectives 6(2), 11–30 (2021).
22. Khalil, S.: Unpacking the link between cousin marriage and women's paid work. Social Science Research 123, 103061 (2024). https://doi.org/10.1016/j.ssresearch.2024.103061.
23. Matsuda, N., Ahmed, T., Nomura, S.: Labor market analysis using big data: The case of a Pakistani online job portal. World Bank, Washington, DC (2019). https://doi.org/10.1596/1813-9450-9063.
24. Nasir, Z.M.: An analysis of occupational choice in Pakistan: A multinomial approach. Pakistan Development Review, 57–79 (2005).
25. Pakistan Bureau of Statistics: Pakistan Labor Force Survey 2018–19. Available at: https://www.pbs.gov.pk/sites/default/files/labour_force/publications/lfs2018_19/lfs_2018_19_final_report.pdf (2019).
26. Pakistan Bureau of Statistics: Pakistan Labor Force Survey 2020–21. Available at: https://www.pbs.gov.pk/sites/default/files/labour_force/publications/lfs2020_21/LFS_2020-21_Report.pdf (2021).
27. Roomi, M.A., Rehman, S., Henry, C.: Exploring the normative context for women's entrepreneurship in Pakistan: A critical analysis. International Journal of Gender and Entrepreneurship 10(2), 158–180 (2018).
28. Sabir, M., Aftab, Z.: Dynamism in the gender wage gap: Evidence from Pakistan. Pakistan Development Review, 865–882 (2007).

29. Sarwar, A., Imran, M.K.: Exploring women's multi-level career prospects in Pakistan: Barriers, interventions, and outcomes. Frontiers in Psychology 10 (2019).
30. Shah, N., Abbasi, N., Iqbal, A.: Interdistrict and interprovincial differentials in correlates of female labour force. Pakistan Development Review 15(4), 424–445 (1976).
31. Siddiqui, R., Siddiqui, R., Akhtar, M.R.: A decomposition of male-female earnings differentials. Pakistan Development Review, 885–898 (1998).
32. Stier, H., Mandel, H.: Inequality in the family: The institutional aspects of women's earning contribution. Social Science Research 38(3), 594–608 (2009).
33. Turaeva, M.R., Becker, C.M.: Daughters-in-law and domestic violence: Patrilocal marriage in Tajikistan. Feminist Economics 28(4), 60–88 (2022).
34. UN Women: National report on the status of women in Pakistan, 2023. Available at: https://pakistan.unwomen.org/sites/default/files/2023-07/summary_-nrsw-inl_final.pdf (2023).
35. Waqar, S., Hanif, R., Loh, J.: Invisibility not invincibility: Pakistani women and the lack of career ascendance. Gender in Management: An International Journal 36(6), 731–744 (2021).
36. World Bank: World Bank Open Data. Available at: https://data.worldbank.org/ (2023).
37. World Bank: Women, business, and the law 2024 – Pakistan. Available at: https://wbl.worldbank.org/content/dam/documents/wbl/2024/snapshots/Pakistan.pdf (2024).
38. World Economic Forum: Global Gender Gap Report 2024. Available at: https://www3.weforum.org/docs/WEF_GGGR_2024.pdf (2024).
39. Worldwide Independent Network of Market Research: Equal opportunities and rights: A global picture. Available at: https://winmr.com/equal-opportunities-and-rights-a-global-picture/ (2021).
40. Zulfiqar, A., et al.: Homemaker or breadwinner: Labour force participation of Pakistani women. Community, Work and Family, 1–20 (2024).